\documentclass[a4paper,pre,twocolumn]{revtex4}
\usepackage{graphicx}
\usepackage{amsmath}

\DeclareMathOperator*{\erf}{erf}

\newcommand\dx{\mathrm{d}x}
\newcommand\kB{k_\mathrm{B}}
\newcommand\ldisp{\ell_\mathrm{disp}}
\newcommand\lmin{l_\mathrm{min}}
\newcommand\lmax{l_\mathrm{max}}
\newcommand\njump{n_\mathrm{jump}}
\newcommand\nl{n_\mathrm{l}}
\newcommand\Qtot{Q_\mathrm{tot}}
\newcommand\wgap{w_\mathrm{g}}
\newcommand\wlane{w_\mathrm{l}}
\newcommand\mum{$\mu$m}

\renewcommand\eqref[1]{Eq.~(\ref{#1})}
\newcommand\figref[1]{Fig.~\ref{#1}}

\begin{document}
\newcommand\micdtu{%
  MIC -- Department of Micro and Nanotechnology,
  Technical University of Denmark\\
  DTU Building~345~East, DK-2800 Kongens Lyngby, Denmark}
\title{%
  A theoretical analysis of the resolution due to diffusion and size-dispersion
  of particles in deterministic lateral displacement devices}
\author{%
  \firstname{Martin}
  \surname{Heller}}
  \homepage{http://www.mic.dtu.dk/microfluidics}
  \affiliation{\micdtu}
\author{%
  \firstname{Henrik}
  \surname{Bruus}}
  \homepage{http://www.mic.dtu.dk/microfluidics}
  \affiliation{\micdtu}
\date{2 November 2007}

\begin{abstract}
We present a model including diffusion and particle-size
dispersion for separation of particles in deterministic lateral
displacement devices also known as bumper arrays. We determine the
upper critical diameter for diffusion-dominated motion and the
lower critical diameter for pure convection-induced displacement.
Our model explains the systematic deviation, observed for small
particles in several experiments, from the critical diameter for
separation given by simple laminar flow considerations.
\end{abstract}

\maketitle
\section{Introduction}
In 2004 \citeauthor{Huang2004}~\cite{Huang2004} developed the
elegant method of particle separation by deterministic lateral
displacement in so-called microfluidic bumper arrays. The method,
which relies on the laminar flow properties characteristic of
microfluidcs, shows a great potential for fast and accurate
separation of particles on the micrometer
scale~\cite{Huang2004,Zheng2005,Inglis2006,Davis2006}. Among the
key assets of the deterministic lateral displacement separation
principle are that clogging can be avoided because particles much
smaller than the feature size of the devices can be separated,
that the devices are passive, i.e.\ the particles bump into solid
obstacles or bumpers, and that the separation process is
continuous.

More precisely, particle transport in microfluidic bumper arrays
is primarily governed by convection due to the fluid flow and
displacement due to interaction with the bumpers in the
array~\cite{Huang2004}. These processes are deterministic and the
critical diameter for separation of relatively large particles in
these devices is well understood in terms of the width of flow
lanes bifurcating around the bumpers in the periodic
arrays~\cite{Inglis2006}. However, if bumper arrays and particles
are scaled down, diffusion will influence the separation process
and affect the critical particle size significantly. Previously
reported data on separation of particles in bumper arrays all show
a bias towards larger critical particle size than that given by
the width of the flow lanes nearest to the bumpers of the
array~\cite{Huang2004, Zheng2005, Inglis2006}. In this work we
extend existing models by adding diffusion and taking
particle-diameter dispersion into account, and thereby explain the
observed discrepancy.

\begin{figure}
\includegraphics{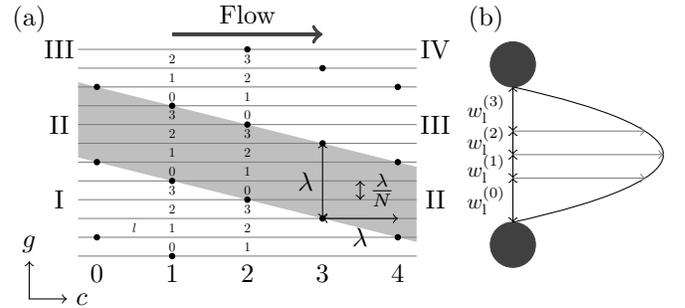}
\caption{(a)~An array of bumpers (black dots) with the definition
of the lane number~$l$ (small arabic numbers), the column
number~$c$ (large arabic numbers), and the gap number~$g$ (roman
numbers). The shaded region illustrates how the shift in the
position of gap~II follows the geometry of the array. (b)~Close-up
of a single gap between two bumpers (disks) in the array. Each of
the four flow lanes carries the same flow rate. Due to the
parabolic flow profile in the gap region, the width $\wlane^{(l)}$
of flow-lane $l$ depends on the position in the
gap.}\label{fig:bumper-view}
\end{figure}

In bumper arrays particles are convected by the fluid flow through
an array of bumpers placed in columns separated by the distance
$\lambda$ in the flow direction, see \figref{fig:bumper-view}(a).
For a given integer $N$, The array is made $N$-periodic in the
flow direction by displacing the bumpers in a given column a
distance $\lambda/N$ perpendicular to the flow direction with
respect to the bumper positions in the previous column. Due to
this periodicity of the array and the laminarity of the flow, the
stream can naturally be divided into $N$ lanes, each carrying the
same amount of fluid flux, and each having a specific path through
the device, see Ref.~\cite{Huang2004}.

For a given steady pressure drop, the fluid in the device moves
with an average velocity~$u_0$. Assuming a parabolic velocity
profile $u(x)$ in the gap of width $\wgap$ between two neighboring
bumpers, see \figref{fig:bumper-view}(b),
\begin{equation}
  u(x) = 6u_0\frac{x}{\wgap}\left(1-\frac{x}{\wgap}\right),
\end{equation}
the total flow rate $\Qtot$ is given by
\begin{equation}
  \Qtot = \int_0^{\wgap}\!u(x)\,\dx = \wgap u_0.
\end{equation}
For an $N$-periodic array, the $N$ flow lanes in a given gap carry
the same flow rate $\Qtot/N$. The width $\wlane^{(l)}$ of lane~$l$
is found by solving
 \begin{equation}
 \frac{\Qtot}{N} = \int_{x^{(l)}}^{x^{(l)}+\wlane^{(l)}} u(x)\,\dx,
 \end{equation}
where $x^{(l)} = \sum_{j = 0}^{l}\wlane^{(j)}$ is the starting
position of lane $l$. In the simple bifurcating flow-lane model
\cite{Huang2004,Inglis2006} the critical diameter $d_\mathrm{c}$
is given as $d_\mathrm{c}/2 = \wlane^{(1)}$. A small particle with
$d<d_\mathrm{c}$ will never leave its initial flow lane and will
thus be convected in the general flow direction following a
so-called zigzag path. Large particles with $d>d_\mathrm{c}$ will
quickly bump against a bumper and from then on be forced by
consecutive bumping to follow the skew direction of the array
geometry, the so-called displacement path. When a particle gets
bumped by a bumper in the array it will be displaced perpendicular
to the flow direction until its center is located one particle
radius~$d/2$ from the surface of the bumper. This corresponds to
$\nl$ lanes of displacement,
\begin{equation}
  \nl = \frac{N}{\wgap u_0}\int_0^{d/2}\!u(x)\,\dx
      = N \frac{d^2}{4\wgap^2}\left(3-\frac{d}{w}\right).
\end{equation}

In the bulk fluid, where the lanes are assumed to have equal width
$\lambda/N$, see \figref{fig:bumper-view}(a), the displaced distance $\ldisp$ is therefore
\begin{equation}
  \label{eq:ldisp}
  \ldisp(d) = \nl\frac{\lambda}{N}
    = \lambda\frac{d^2}{4\wgap^2}\left(3-\frac{d}{\wgap}\right).
\end{equation}

In this work we extend the simple bifurcating flow-lane model by
including diffusion and particle-diameter dispersion.

\section{Model including diffusion}
During the average time $\tau= \lambda/u_0$ it takes a particle to
move by convection from one column to the next, it also diffuses.
We assume that the diffusion process perpendicular to the flow
direction is normally distributed with mean value zero and
variance
 \begin{equation}
 \sigma^2=2D\tau,
 \end{equation}
where the diffusivity~$D$ is given by the Stokes--Einstein
expression
 \begin{equation}
  D = \frac{\kB T}{3\pi\eta d}.
 \end{equation}
Here $\kB$ is Boltzmann's constant, $T$~is the temperature and
$\eta$~is the viscosity of the fluid.

\subsection{Diffusion model}
\begin{figure}
\includegraphics{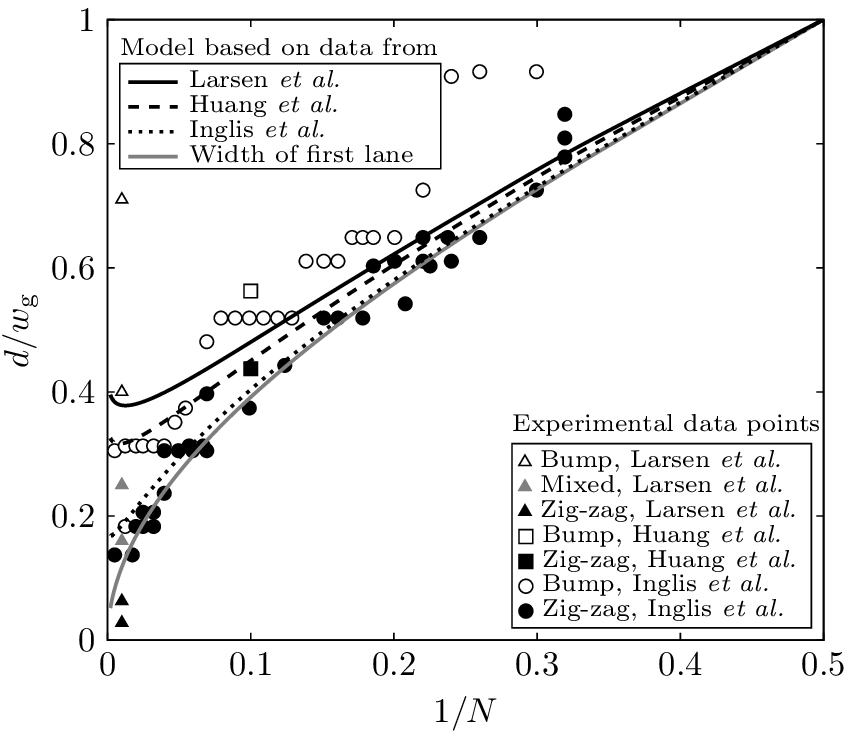}
\caption{Our model applied to the experimental data of
\citeauthor{Inglis2006}~\cite{Inglis2006},
\citeauthor{Huang2004}~\cite{Huang2004} and
\citeauthor{Larsen2006}~\cite{Larsen2006}. Particle diameter $d$
over gap width $\wgap$ is plotted as a function of the inverse
period $1/N$. The black lines show the theoretically predicted
critical particle size for the bumper arrays used by
\citeauthor{Inglis2006}~\cite{Inglis2006},
\citeauthor{Huang2004}~\cite{Huang2004} and
\citeauthor{Larsen2006}~\cite{Larsen2006}, respectively. The
experimental data points are representing particles following
bumper  paths (open symbols), zigzag paths (solid black symbols),
and  neither of these paths (solid gray
symbols).}\label{fig:inglisdata}
\end{figure}

In order to escape bumping, a particle must in the time interval
$\tau$ diffuse more than the difference $\ldisp-\lambda/N$ between
the bulk displacement and the shift in position of the next bumper.
The probability $p_\mathrm{esc}$ for this to happen is given
in terms of the error function as
 \begin{equation}\label{eq:p_esc}
  p_\mathrm{esc}(d) = \frac{1}{2} - \frac{1}{2}\erf\mathopen{}\left(
    \frac{\ldisp(d)-\frac{1}{N}\,\lambda}{\sqrt{2}\,\sigma(d)}
  \right),
 \end{equation}
where we have introduced the $d$-dependence explicitly. When a
particle is transported through a bumper array it must bump at
every bumper within one period of the array in order to be
displaced one gap at the outlet. Thus, if the particle escapes at
least one time in $N$ attempts, it will not be displaced. We
define the critical particle size $d^{{}}_\mathrm{c}$ to be the
size for which half of the particles escape bumping as they pass
one period of the array. Thus $d^{{}}_\mathrm{c}$ can be found by
solving
 \begin{equation}
  \sum_{k=1}^N
    \begin{pmatrix}
      N \\ k
    \end{pmatrix}
    p_\mathrm{esc}(d_\mathrm{c})^k
    \Bigl[1-p_\mathrm{esc}(d_\mathrm{c})\Bigr]^{N-k}
    = \frac{1}{2}.
 \end{equation}
In \figref{fig:inglisdata} we have plotted the critical particle
size as a function of the bumper period for parameter values
corresponding to the bumper arrays used by
\citeauthor{Huang2004}~\cite{Huang2004} (dashed line),
\citeauthor{Inglis2006}~\cite{Inglis2006} (dotted line), and
\citeauthor{Larsen2006}~\cite{Larsen2006} (full line).

\subsection{Comparison to experiments}
The observation that the critical particle size in a bumper device
is larger than the width of the first flow lane is also supported
by the experimental data in Fig.~2 of Ref.~\cite{Inglis2006}. Our
model suggests that the deviation of the critical particle size
from the width of the first flow lane can be explained by
diffusion of the particles. In \figref{fig:inglisdata} above, it
is seen how well the lines are predicting the transition between
zigzag paths and displacement paths: the thick line is dividing
open and closed triangles, the dashed line is dividing open and
closed squares, and, to a lesser extent, the dotted line is
dividing open and closed circles.

Using parameter values corresponding to the bumper device
presented by \citeauthor{Huang2004}~\cite{Huang2004} our model
predicts a critical particle diameter of $0.45$ times the width of
the gap for the particles traveling through the device with an
average velocity of \mbox{$400$\,\mum/s}. This is in good
agreement with Fig.~2(a) in Ref.~\cite{Huang2004}.

\section{A discrete model including diffusion and dispersion}
The particles used in the experiments on separation of particles
are not mono-disperse. Their average diameters are distributed
around a certain mean value with a relative standard deviation
$\Delta d/d$, which typically is $20\,\%$, $10\,\%$ and~$5\,\%$
for particles with \mbox{$d=25$\,nm}, \mbox{$d=100$\,nm}
and~\mbox{$d=500$\,nm}, respectively.

In the following we introduce a discrete model of the transport of
particles with different diameters $d$ in an $N$-periodic bumper
array taking convection, diffusion and size-dispersion into
account. The model allows us to study the relative influence of
all three phenomena on the separation efficiency in a fast and
simple manner. In particular our results suggest that the critical
size for separation or displacement, studied above, must be
supplemented by a smaller critical size below which pure diffusion
governs the motion of the particles in the bumper array. This
prediction has not yet been tested experimentally.

\subsection{Definition of the discrete model}
At any instant, a particle is assumed to be positioned in the
center of a flow lane $l$ situated in a gap $g$ in some column $c$
of the array. For simplicity we further assume that the size
distribution of any given set of particles is a normal
distribution with a mean value given by the size quoted by the
manufacturer and a relative standard deviation of $10\,\%$.

By convection any given particle moves from one column to the
next. If it ends up in the last lane ($l=N-1$) in one gap, it will
be shifted to the first lane ($l=0$) in the subsequent gap.
Otherwise it will stay in the current gap and move up one lane. In
our model pure convection is therefore described by the discrete
map
\begin{equation}\label{eq:convection}
  (c,g,l) \mapsto
  \begin{cases}
    (c+1,g+1,0) & \text{, if $l = N-1$,} \\[1mm]
    (c+1,g,l+1) & \text{, otherwise.}
  \end{cases}
\end{equation}

Because of the finite diameter~$d$ of the particle there is a
minimum and a maximum lane number that it can occupy. The minimum
lane number is the smallest integer $\lmin$ that satisfies
\begin{subequations}
  \begin{align}
    \sum_{l=0}^{\lmin}\wlane^{(l)} &> \frac{d}{2}.
\intertext{%
Similarly, the maximum lane number $\lmax$ is the largest integer that satisfies}
    \sum_{l=\lmax}^{N-1} \wlane^{(l)} &> \frac{d}{2}.
  \end{align}
\end{subequations}
Consequently, the simple convection mapping from
\eqref{eq:convection} needs to be modified to account for the
finite size of the particles
\begin{equation}\label{eq:displacement}
   (c,g,l) \mapsto
   \begin{cases}
     (c+1,g+1,{\lmin}) & \text{, if $l = N-1$,}     \\[1mm]
     (c+1,g,l+1)       & \text{, if $l<{\lmax}-1$,} \\[1mm]
     (c+1,g,{\lmax})   & \text{, otherwise.}
   \end{cases}
\end{equation}
The above convection scheme accounts for the separation of
particles in deterministic lateral displacement devices according
to size. The critical particle radius predicted by this model is
 \begin{equation}\label{eq:dcDef}
 d_\mathrm{c} = 2\wlane^{(0)}
 \end{equation}
in accordance with the geometric arguments of
Ref.~\cite{Inglis2006}.

\begin{figure*}
\includegraphics{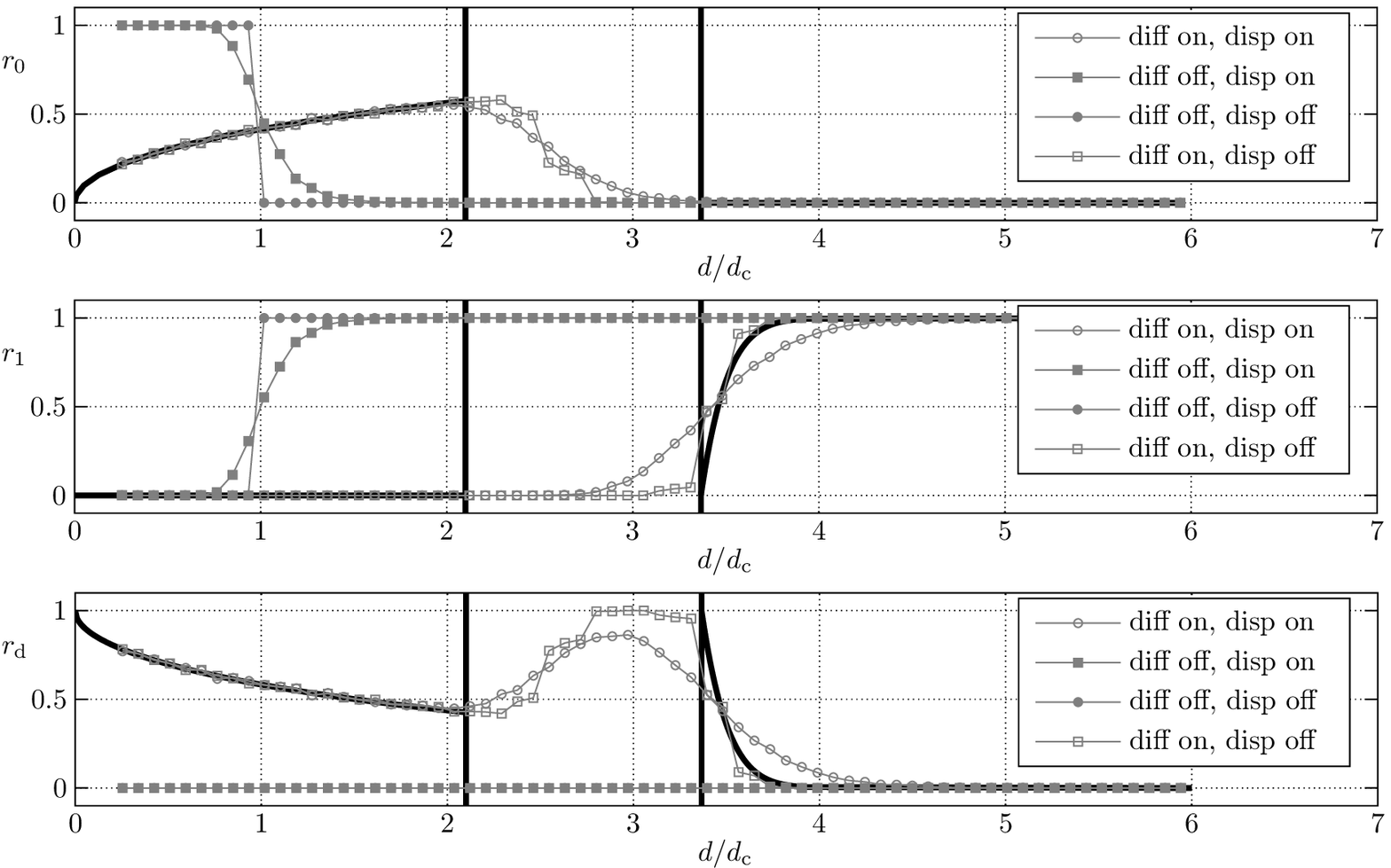}
\caption{Relative number $r_0$, $r_1$ and $r_\mathrm{d}$ of
particles following the zigzag path, the displacement path, and
neither of these two paths, respectively, plotted as a function of
the normalized, average particle diameter $d/d_\mathrm{c}$. The
parameters of the bumper array is taken form
Ref.~\cite{Larsen2006}: $N=100$, \mbox{$\lambda=8$\,\mum},
\mbox{$\wgap=1$\,\mum}, \mbox{$L=20N\lambda=16$\,mm}, and
\mbox{$u_0=250$\,\mum/s}. The buffer liquid is water at room
temperature. Neglecting diffusion (solid symbols), the particles
follow the zigzag path if $d < d_\mathrm{c} = 118$\,nm. If
$d>d_\mathrm{c}$ they follow the displacement path. Including
diffusion (open symbols), the small particles are not influenced
by the bumpers. For $d \gtrsim 2.1\, d_\mathrm{c}$ the influence
of the bumpers set in, and for $d \gtrsim3.4\, d_\mathrm{c}$ the
particles follow the displacement path. The thick black vertical
lines indicate the particle size when small particles stop
behaving purely diffusive (left-most lines) and when large
particles begin a purely deterministic displacement (right-most
lines).}
  \label{fig:weights}
\end{figure*}

To characterize the separation quantitatively, we define the
relative particle numbers $r_0$, $r_1$ and $r_\mathrm{d}$ as
\begin{subequations}
  \newlength\equationwidth\setlength\equationwidth{6.42cm}
  \begin{align}
    r_0 &= \parbox[t]{\equationwidth}{relative number of particles following the zigzag path.}\\
    r_1 &= \parbox[t]{\equationwidth}{relative number of particles following the displacement path.}\\
    r_\mathrm{d} &= \parbox[t]{\equationwidth}{relative number of all other particles.}
  \end{align}
\end{subequations}
With these definitions the sum $r_0+r_1+r_\mathrm{d}$ is always
unity. If $r_0=1$ all particles follow the zigzag path and if
$r_1=1$ all particles follow the displacement path. If
$r_\mathrm{d}\neq0$ some of the particles end up at positions not
explained by the deterministic analysis of the separation process.
In~\figref{fig:weights} we have plotted the relative particle
numbers $r_0$, $r_1$, and $r_\mathrm{d}$ as a function of average
particle diameter $d$.

\subsection{Pure mono-disperse convection}
For mono-disperse and non-diffusing particles, the model results,
as expected, in two modes: the zigzag mode and the displacement
mode, see the closed circles in \figref{fig:weights}. For
$d<d_\mathrm{c}$ we have $r_0=1$, and for $d\geq d_\mathrm{c}$ we
have $r_1=1$, while we always have $r_\mathrm{d}=0$. The relative
particle numbers can therefore be written as
 \begin{equation}
  \left(r_0,r_\mathrm{d},r_1\right) =
  \begin{cases}
    (1,0,0) & \text{for $d<d_\mathrm{c}$} \\[1mm]
    (0,0,1) & \text{for $d \geq d_\mathrm{c}$}
  \end{cases}.
 \end{equation}

\subsection{Influence of size dispersion}
If we assume that the particles are not mono-disperse, but are
distributed around a mean size $d$ with standard deviation $\Delta
d$, the shift as a function of $d$ from the zigzag mode to the
displacement mode happens gradually instead of abruptly at a
certain critical size $d_\mathrm{c}$ (\figref{fig:weights}, closed
squares). The relative number of particles following the zigzag
path $r_0$ can be found by integrating over all particle sizes
smaller than the critical diameter given by the array geometry
\begin{subequations}
  \begin{align}
    r_0 &= \int_{-\infty}^{d_\mathrm{c}}\!
             \frac{1}{\sqrt{2\pi(\Delta d)^2}}
             \exp\mathopen{}\left(
               \frac{-(s-d)^2}{2(\Delta d)^2}
             \right)\,\mathrm{d}s.
  \end{align}
Similarly, the relative number of particles following the
displacement path $r_1$ can be found by integrating over all
particle sizes larger than $d_\mathrm{c}$
  \begin{align}
    r_1 &= \int_{d_\mathrm{c}}^{\infty}\!
             \frac{1}{\sqrt{2\pi(\Delta d)^2}}
             \exp\mathopen{}\left(
               \frac{-(s-d)^2}{2(\Delta d)^2}
             \right)\,\mathrm{d}s.
  \end{align}
\end{subequations}
The system is still a bi-modal system because $r_\mathrm{d}=0$ for all particle sizes.

\subsection{Influence of diffusion}
In 1D during the time step $\tau$ a particle diffuses the distance
$\ell$, the average of which is the size-dependent diffusion
length $\sigma(d)$ given by
 \begin{equation}
 \label{eq:ldiff}
  \sigma(d) = \langle \ell \rangle = \sqrt{2D\tau}
         = \sqrt{2\frac{\kB T}{3\pi\eta d}
             \frac{\lambda}{u_0}}.
 \end{equation}
In our model we discretize the transverse diffusion as the
properly rounded number of flow lanes crossed by the particle
during diffusion,
 \begin{equation}
 \njump= \frac{N}{\lambda}\:\ell.
 \end{equation}
The addition of diffusion smears out the displacement of the
particles and causes the critical radius to be larger than in the
diffusion-less case (\figref{fig:weights}, open symbols).

\subsubsection{Bumping criterion for small particles}
Very small particles are completely dominated by diffusion, and
the particle distribution at the end of the array is simply given
by the transverse diffusion of the particles during the time
$L/u_0$ it takes for the particle to be convected all the way $L$
through the array. For small particles we therefore have
\begin{equation}
  r_0 = \int_{-\frac32\frac{\lambda}{N}}
            ^{\frac32\frac{\lambda}{N}}\!
          \frac{1}{\sqrt{2\pi\sigma^2}}
          \exp\mathopen{}\left(
            \frac{-x^2}{2\sigma^2}
          \right)\,\dx,
\end{equation}
where $\sigma^2 = 2D L/u_0$.

As the particle diameter $d$ is increased, the bumpers begin to
become important as the diffusion length $\sigma(d)$,
Eq.~(\ref{eq:ldiff}), is decreased and becomes equal to the
displacement length $\ldisp$, Eq.~(\ref{eq:ldisp}). Using the
criterion $\sigma(d_1) = \ldisp$, with the parameter values used
in Fig.~\ref{fig:weights}, we find that particles stop behaving as
small diffusion-dominated particles and start interacting with the
bumpers when $d_1=2.1\, d_\mathrm{c}$. This cross-over value is
indicated by the left-most vertical lines in
Fig.~\ref{fig:weights}, and it fits well with the simulation data.

\subsubsection{Bumping criterion for large particles}
Large particles will interact with the bumpers at every row in the
array and their position is thus reset at every bump to be
$\ldisp$. Diffusion can therefore be neglected for such particles,
they all follow the displacement path, and $r_1=1$.

As the particle diameter is lowered, the probability
$p_\text{esc}$ that a particle escapes the displacement path can
be estimated by the probability of diffusing from the displaced
position to the last flow lane in the gap, i.e.~the distance
$\ldisp-\frac{\lambda}{N}$. This probability $p_\mathrm{esc}$ is
given by \eqref{eq:p_esc}.

In order to follow the displacement path, a particle must bump at
each row in the array. If we consider a $N$-periodic array with
$m$ full periods, the particles will have $mN$ bumping
opportunities as they pass through the entire array. If a particle
evades bumping at a bumper, it will be convected by the flow
through a full period of the array before bumping is possible
again. Because of the escape, it will miss $N$ bumping
opportunities and end up one gap from the displacement path.
Consequently, if a particle escapes one time, it will only have
$(m-1)N$ bumping opportunities and has therefore escaped bumping
with a probability of $1/[(m-1)N]$. The upper critical particle
size $d_2$ for convection-induced displacement is defined using
\eqref{eq:p_esc} as
\begin{equation}
  p_\mathrm{esc}(d_2) =  \frac{1}{(m-1)N}.
\end{equation}
For the device used in the experiments by
\citeauthor{Larsen2006}~\cite{Larsen2006} we find
$d_2=3.4d_\mathrm{c}$. The predicted upper limit for diffusion
dominated motion and the lower limit for convection-induced
displacement compares well with the experimental observation by
\citeauthor{Larsen2006}~\cite{Larsen2006} that some particles end
up between the displacement path and the zigzag path
(Fig.~\ref{fig:inglisdata}, gray triangles).

\section{Conclusion}
Experimental data on separation of particles in bumper arrays all
show a systematic deviation from predictions made from the
bifurcating flow lane
model~\cite{Inglis2006,Huang2004,Larsen2006}. By adding diffusion
to the bifurcating flow lane model, we explain the discrepancy. In
addition, we have suggested a simple discrete model for simulating
particle separation in bumper arrays. The presented model takes
particle diffusion and size dispersion into account and has been
validated against experimental data for a bumper device with
period $N=100$. The transition from zigzag path to displacement
path happens at a particle size $2.1$ to $3.4$ times larger than
the critical particle size predicted from geometrical arguments.
This is in correspondence with the experimental observations from
Larsen~\cite{Larsen2006}. Our discrete model and the estimates
presented in this paper suggest that particles of twice the size
of the geometrical critical size of the $N=100$ bumper device
behave diffusively and are not affected by the bumpers because the
small diffusive particles rarely come into contact with the
bumpers due to random Browninan motion.

\section{Acknowledgement}
We thank Asger Vig Larsen, Anders Kristensen,  Jason Beech, and
Jonas Tegenfeldt for inspiring discussions on experimental issues.
Martin Heller was supported by the Danish Research Council for
Technology and Production Sciences grant no.~26-04-0074.


\end{document}